\documentclass[%
floatfix,
showkeys,
nofootinbib, %
superscriptaddress, %
]{revtex4-1}

\usepackage{cmap}

\usepackage{ucs}
\usepackage[utf8x]{inputenc}
\usepackage[T1,T2A]{fontenc}
\usepackage[german,russian,english]{babel}

\usepackage[sort&compress]{natbib}
\usepackage{amsmath}
\usepackage{amssymb}

\usepackage{mathtools}
\mathtoolsset{
showonlyrefs,
mathic = true
}

\allowdisplaybreaks

\usepackage{hyperref}
\hypersetup{backref,
 colorlinks=false}
\hypersetup{pdfborder=0 0 0}

\usepackage{microtype}
\UseMicrotypeSet[protrusion]{alltext}

\usepackage{graphicx}

\usepackage[scanall]{psfrag}

\usepackage{listings}
\usepackage{listingsutf8}
\usepackage{fixltx2e}

\usepackage{nicefrac}

\makeatletter
\def\ps@pprintTitle{%
     \let\@oddhead\@empty
     \let\@evenhead\@empty
     \let\@oddfoot\@empty
     \let\@evenfoot\@oddfoot}
\makeatother

\usepackage{physics}

\begin{document}

\title{Constrained Hamiltonian approach to the Maxwell theory}

\author{Dmitry S. Kulyabov}
\email{kulyabov-ds@rudn.ru}
\affiliation{Department of Applied Probability and Informatics,\\
  Peoples' Friendship University of Russia (RUDN University),\\
  6 Miklukho-Maklaya St, Moscow, 117198, Russian Federation}
\affiliation{Laboratory of Information Technologies\\
  Joint Institute for Nuclear Research\\
  6 Joliot-Curie, Dubna, Moscow region, 141980, Russia}

\author{Anna V. Korolkova}
\email{korolkova-av@rudn.ru}
\affiliation{Department of Applied Probability and Informatics,\\
  Peoples' Friendship University of Russia (RUDN University),\\
  6 Miklukho-Maklaya St, Moscow, 117198, Russian Federation}

\author{Migran N. Gevorkyan}
\email{gevorkyan-mn@rudn.ru}
\affiliation{Department of Applied Probability and Informatics,\\
  Peoples' Friendship University of Russia (RUDN University),\\
  6 Miklukho-Maklaya St, Moscow, 117198, Russian Federation}

\author{Leonid A. Sevastianov}
\email{sevastianov-la@rudn.ru}
\affiliation{Department of Applied Probability and Informatics,\\
  Peoples' Friendship University of Russia (RUDN University),\\
  6 Miklukho-Maklaya St, Moscow, 117198, Russian Federation}
\affiliation{Bogoliubov Laboratory of Theoretical Physics\\
  Joint Institute for Nuclear Research\\
  6 Joliot-Curie, Dubna, Moscow region, 141980, Russia}

\begin{abstract}
The most common physical formalisms are the Lagrangian formalism and
the Hamiltonian formalism. From the superficial point of view, they are
one and the same, but rewritten in other terms. However, it seems that
the Hamiltonian formalism has a richer structure and is more
convenient for studying the electromagnetic field, especially in the
formalization of its geometrization. Unfortunately for field problems,
there is a whole set of Hamiltonian formalisms. The authors study the
applicability of different variants of the Hamiltonian formalism to
the problems of electrodynamics. In this paper we consider the
Hamiltonian formalism with constraints.
\end{abstract}


\maketitle

\section{Introduction}
\label{sec:intro}

In the study of electromagnetic and optical phenomena Hamiltonian formalism is often used. The main drawback of Hamiltonian formalism
it seems to be poorly developed for field systems. For
descriptions of field systems, we propose several variants of the Hamiltonian
formalism. Previously, we considered the possibility of building
symplectic~\cite{kulyabov:2017:sfm:instantaneous_hamiltonian} and
multipulse~\cite{kulyabov:2017:sfm:geometric_lagrangian}
Hamiltonian formalism. In this paper we consider the construction of a Hamiltonian formalism with constraints.

\section{Notations and conventions}
\label{sec:notation}

  \begin{enumerate}

  \item We will adhere to the following agreements. Greek indices
    ($\alpha$, $\beta$) will refer to the four-dimensional space.
    Latin
    indices from the middle of the alphabet ($i$, $j$, $k$) will refer
    to the three-dimensional space.

  \item In the theoretical description, Latin indices will refer to
    the space of arbitrary dimension.

  \item The comma in the index denotes a partial derivative with respect to
    corresponding coordinate ($f_{, i} := \partial_{i}f$);  the semicolon
    denotes a covariant derivative ($f_{;i} := \nabla_{i} f$).

  \item The CGS symmetrical system~\cite{sivukhin:1979:ufn::en} is used for
    notation of the equations of electrodynamics.

  \end{enumerate}

\section{Hamiltonian formalism}
\label{sec:hamilton}

There are several variants of the Hamiltonian formalism.
\begin{itemize}
\item The symplectic Hamiltonian formalism~\cite{vilasi:book:hamiltonian::en,kulyabov:2017:sfm:instantaneous_hamiltonian}.
\item The Dirac–Bergman Hamiltonian formalism for systems with constraints~\cite{dirac:lectures-quantum-mechanics::en, bergman:1955:lagrangians}.
\item The Hamilton–De Donder Hamiltonian formalism~\cite{sardanashvily:1995:hamiltonian_formalism}.
\item The multimomentum Hamiltonian formalism~\cite{sardanashvily:1995:hamiltonian_formalism,
    giachetta:1999:covariant_hamilton, giachetta:2009:advanced_classical}.
\end{itemize}

Consider major points of the Hamiltonian formalism.

Let the system be described by some quantity called the action:
\begin{equation}
S[q^i] = \int \dd^4 \mathcal{L}(x^i, q^i, \Dot{q}^i).
\end{equation}
Lagrangian (Lagrangian density)
$(x^I, q^i, \Dot{q}^i)$ depends on the generalized coordinates $q$ and their
first derivatives of $\Dot{q}$.

In the transition to the Hamiltonian formalism
the system is described by generalized coordinates
$q^I$ and generalized momentum:
\begin{equation}
  \label{eq:hamilton:p-L}
p_i = \pdv{\mathcal{L}}{\Dot{q}^i}.
\end{equation}

One can construct the Legendre transform in velocities:
\begin{equation}
\mathcal{H}(q^i, p_i) := p_i \Dot{q}^i − \mathcal{L}(q^i, \Dot{q}^i).
\end{equation}
The function $\mathcal{H}$ is called a Hamiltonian (Hamiltonian
density). The Hamiltonian of the system depends only on
generalized coordinates and momenta.

Using the connection between Hamiltonian and Lagrangian, we consider the action:
\begin{equation}
S[q^i,p_i] = \int \dd^4 \qty(p_i\Dot{q}^i - \mathcal{H}(q^i, p_i)).
\end{equation}
The corresponding system of Euler--Lagrange equations has the form:
\begin{equation}
  \label{eq:hamilton:ham-eq-S}
  \begin{gathered}
    \fdv{S}{q^i} = - \Dot{p}_i - \fdv{\mathcal{H}}{q^i} = 0,
    \\
    \fdv{S}{p_i} = \Dot{q}^i - \fdv{\mathcal{H}}{p_i} = 0.
  \end{gathered}
\end{equation}

Let us define the Poisson brackets:
\begin{equation}
[f, g] := \pdv{f}{q^i} \pdv{g}{p_i} - \pdv{f}{p_i} \pdv{g}{q^i}.
\end{equation}

Then we can rewrite equation~\eqref{eq:hamilton:ham-eq-S} as:
\begin{equation}
  \begin{gathered}
    \Dot{q}^i = \qty[q^i, \mathcal{H}] = \fdv{\mathcal{H}}{p_i},
    \\
    \Dot{p}_i = \qty[p_i, \mathcal{H}] = - \fdv{\mathcal{H}}{q^i}.
  \end{gathered}
\end{equation}

\section{Hamiltonian dynamics with constraints}
\label{sec:constrain}

If the Lagrangian is singular by velocities:
\begin{equation}
\det \qty[\pdv{\mathcal{L}}{\Dot{q}^i}{\Dot{q}^j}] = 0,
\end{equation}
it is not possible to express all momentum by
formula~\eqref{eq:hamilton:p-L}. In this case
one gets only possible momentum and for the rest
the concept of constraints~\cite{dirac:lectures-quantum-mechanics::en} is used.

We will consider the system with Lagrangian
$\mathcal{L} (x^k,q^k,p_k)$, $I=\overline{1,n}$. Also consider
set of constraints:
\begin{equation}
  \varphi^a (x^k,q^k,\Dot{q}^k), \quad a=\overline{1,m},
  \quad m \leqslant n.
\end{equation}

Action minimum, in case the trajectory satisfy the equations of connection,
is interpret also on trajectories without constraints, but with Lagrangian with
constraints: 
\begin{equation}
  L(x^k,q^k,\Dot{q}^k) =
  \mathcal{L}(x^k,q^k,\Dot{q}^k) -
  \lambda_a (x^k,q^k,\Dot{q}^k) \varphi^a (x^k,q^k,\Dot{q}^k).
\end{equation}

Hamilton's equations take the following form:
\begin{equation}
  \begin{gathered}
    \Dot{q}^i = \fdv{\mathcal{H}}{p_i} + \lambda_a \pdv{\varphi^a}{p_i},
    \\
    \Dot{p}_i = - \fdv{\mathcal{H}}{q^i} - \lambda_a \pdv{\varphi^a}{q^i}.
  \end{gathered}
\end{equation}

Lagrange multipliers are found from the condition of preserving constraints:
\begin{equation}
\Dot{\varphi^a} = [\varphi^a, \mathcal{H}] + \lambda_a \varphi^a = 0.
\end{equation}

\section{The Hamiltonian of the electromagnetic field with constraints}
\label{sec:em-hamilton}

We consider the construction of a Hamiltonian formalism with constraints for the case
electromagnetic field.

Write the Lagrangian of the electromagnetic field~\cite{stratton:1948::en}:
\begin{equation}
\mathcal{L} \qty(x^{\alpha}, A_{\beta}, A_{\alpha, \beta}) 
= 
- \frac{1}{16\pi c} F_{\alpha\beta} F^{\alpha\beta} 
\sqrt{-g}
-
\frac{1}{c^2} A_\alpha j^\alpha 
\sqrt{-g}.
\end{equation}

Since $F_{00}=0$,
\begin{equation}
\pdv[2]{\mathcal{L}}{\Dot{A}_0}=0.
\end{equation}
That is, the Lagrangian is irregular. Therefore, we need to find
constraints for the case $p^0$. Because
\begin{equation}
  p^0 = \pdv{\mathcal{L}}{\Dot{A}_0} = 0,
\end{equation}
then enter the constraints
\begin{equation}
  \label{eq:em-hamilton:varphi}
  \varphi := p^0 \approx 0.
\end{equation}
The $\approx$ symbol indicates that the equality must be performed on
surfaces of all constraints.

Let's write down expressions for momentum:
\begin{equation}
  p_i = \pdv{\mathcal{L}}{\Dot{A}^i} =
  \frac{1}{c} \pdv{\mathcal{L}}{A^i_{,0}}
  =
  -
  \frac{\sqrt{-g}}{4\pi c^2}
  \qty[
  A_{i,0} - A_{0,i}
  ]
  .
\end{equation}

Let us express time derivatives of $A_i$ by momentum:
\begin{equation}
\Dot{A}_i = c A_{i,0} = - \frac{4\pi c^3}{\sqrt{-g}} p_i + c A_{0,i}.
\end{equation}

Let us construct the Hamiltonian
\begin{equation}
  H =
  p^i \Dot{A}_i - \mathcal{L}
  +
  \lambda p^0
  =
  p^{i}\qty(-\frac{4\pi c^3}{\sqrt{-g}} p_i + c A_{0,i})
  +
  \frac{\sqrt{-g}}{16\pi c} F_{\alpha\beta} F^{\alpha\beta} 
  +
  \frac{\sqrt{-g}}{c^2} A_\alpha j^\alpha
  +
  \lambda p^0
\end{equation}

Hamilton's equations take the following form:
\begin{gather}
  \Dot{A}_0 = \pdv{H}{p^0} = \lambda,
  \label{eq:em-hamilton:dotA0}
  \\
  \Dot{A}_{i} = \pdv{H}{p^{i}} = 
  -
  \frac{4\pi c^3}{\sqrt{-g}} p_i
  +
  c A_{0,i}
  ,
  \label{eq:em-hamilton:dotAi}
  \\
  \Dot{p}^0 = -\fdv{H}{A_0} =
  - \pdv{H}{A_0}
  + \partial_{\alpha} \pdv{H}{A_{0,\alpha}}
  =
  -c p^{i}_{,i} + \frac{\sqrt{-g}}{c^2} j^0,
  \label{eq:em-hamilton:dotp0}
  \\
  \Dot{p}^i = - \fdv{H}{A_i} = - \pdv{H}{A_i}
  + \partial_{\alpha} \pdv{H}{A_{i,\alpha}}
  =
  - \frac{\sqrt{-g}}{c^2} j^{i}
  +
  \frac{\sqrt{-g}}{16\pi c} F^{\alpha i}_{,\alpha}
  .
  \label{eq:em-hamilton:dotpi}
\end{gather}

We show that the resulting system of equations is equivalent to Maxwell 
equations (which seems obvious).

From the equation~\eqref{eq:em-hamilton:dotAi} we obtain
\begin{equation}
  \label{eq:em-hamilton:dotAi:pi}
  p_i =
  \frac{\sqrt{-g}}{4\pi c^3}
  \qty(
  c A_{0,i} - \Dot{A}_{i} 
  )
  =
  \frac{\sqrt{-g}}{4\pi c^2}
  \qty(
  A_{0,i} - A_{i,0}
  )
  =
  \frac{\sqrt{-g}}{4\pi c^2}
  F_{i 0}.
\end{equation}
We raise the indices in the equation~\eqref{eq:em-hamilton:dotAi:pi} and substitute
the result in the equation~\eqref{eq:em-hamilton:dotpi}.
Writing the tensor components $F^{{\alpha}{\beta}}$
\begin{equation}
F^{{\alpha}{\beta}}=
\begin{pmatrix}
0 & -{D}^1 & -{D}^2 & -{D}^3 \\
{D}^1 & 0 & -{H}_3 & {H}_2 \\
{D}^2 & {H}_3 & 0 & -{H}_1 \\
{D}^3 & -{H}_2 & {H}_1 & 0
\end{pmatrix},
\label{eq:g^ab}
\end{equation}
we get the equation
\begin{equation}
  \frac{1}{\sqrt{{}^{3}g}}
  \qty[\partial_{{j}}H_{{k}}-\partial_{{k}}H_{{j}}]
  =
  -\frac{1}{c}\partial_t D^{{i}}+\frac{4\pi}{c}j^{{i}}.
\end{equation}

Similarly from~\eqref{eq:em-hamilton:dotp0} taking into account
equation~\eqref{eq:em-hamilton:dotAi:pi} and
connection~\eqref{eq:em-hamilton:varphi}
we obtain another Maxwell equation.
\begin{equation}
\frac{1}{\sqrt{{}^{3}g}}
\partial_{{i}} \qty(\sqrt{{}^{3}g} D^{{i}}) = 4\pi \rho.
\end{equation}

Thus, we showed that the resulting Hamiltonian gives
inhomogeneous Maxwell equations. It is obvious that the homogeneous Maxwell equations
is not obtained from the Hamiltonian and from the Bianchi identities for
tensor $F_{\alpha \beta}$.

\section{Conclusion}
\label{sec:conclusion}

Полученные уравнения
The authors applied the Dirac--Bergman method to the Maxwell equations. Obtained equations
is equivalent to inhomogeneous Maxwell equations. The main feature of our approach is that Maxwell's equations is considered in covariant form and in an arbitrary Riemannian coordinates.

\begin{acknowledgments}

The publication has been prepared with the support of the ``RUDN University Program 5-100''
and funded by Russian Foundation for Basic Research (RFBR) according to the research project
No~19-01-00645.

\end{acknowledgments}

 \bibliographystyle{elsarticle-num}

\bibliography{bib/hamilton-geom/cite}

\end{document}


\title{Гамильтонов подход со связями к теории Максвелла}

\author{Д. С. Кулябов}
\email{kulyabov-ds@rudn.ru}
\affiliation{Кафедра прикладной информатики и теории вероятностей,\\
  Российский университет дружбы народов,\\
  117198, Москва, ул. Миклухо-Маклая, д. 6}
\affiliation{Лаборатория информационных технологий,\\
  Объединённый институт ядерных исследований,\\
  ул. Жолио-Кюри 6, Дубна, Московская область, Россия, 141980}

\author{А. В. Королькова}
\email{korolkova-av@rudn.ru}
\affiliation{Кафедра прикладной информатики и теории вероятностей,\\
  Российский университет дружбы народов,\\
  117198, Москва, ул. Миклухо-Маклая, д. 6}

\author{М. Н. Геворкян}
\email{gevorkyan-mn@rudn.ru}
\affiliation{Кафедра прикладной информатики и теории вероятностей,\\
  Российский университет дружбы народов,\\
  117198, Москва, ул. Миклухо-Маклая, д. 6}

\author{Л. А. Севастьянов}
\email{sevastianov-la@rudn.ru}
\affiliation{Кафедра прикладной информатики и теории вероятностей,\\
  Российский университет дружбы народов,\\
  117198, Москва, ул. Миклухо-Маклая, д. 6}
\affiliation{Лаборатория теоретической физики,\\
  Объединённый институт ядерных исследований,\\
  ул. Жолио-Кюри 6, Дубна, Московская область, Россия, 141980}

\begin{abstract}

Наиболее распространённые физические формализмы: лагранжев формализм и
гамильтонов формализм. С поверхностной точки зрения они суть одно и
тоже, но переписанное в других терминах. Однако представляется, что
гамильтонов формализм имеет более богатую структуру и более удобен для
изучения электромагнитного поля, особенно в формализме его
геометризации. К сожалению для полевых задач существует целый набор
гамильтоновых формализмов. Авторы изучают применимость разных
вариантов гамильтонового формализма к задачам электродинамики. В
данной работе рассматривается гамильтонов формализм со связями.
\end{abstract}

  \keywords{лагранжев формализм, гамильтонов формализм,
    гамильтонов формализм со связями,
    уравнения Максвелла}

\maketitle

\section{Введение}
\label{sec:intro}

При изучении электромагнитных и оптических явлений часто применяется
гамильтоново описание. Основным недостатком гамильтонова формализма
представляется его слабая разработанность для полевых систем. Для
описания полевых систем предложено несколько вариантов гамильтонова
формализма. Ранее мы рассматривали возможность построения
симплектического~\cite{kulyabov:2017:sfm:instantaneous_hamiltonian} и
многоимпульсного~\cite{kulyabov:2017:sfm:geometric_lagrangian}
гамильтоновых формализмов. В
данной работе мы рассматриваем построение гамильтонова формализма со связями.

\section{Обозначения и соглашения}
\label{sec:notation}

  \begin{enumerate}

  \item Будем придерживаться следующих соглашений.  Греческие индексы
    ($\alpha$, $\beta$) будут относиться к четырёхмерному
    пространству. Латинские индексы из середины
    алфавита ($i$, $j$, $k$) будут относиться к трёхмерному
    пространству.

  \item В теоретическом описании латинские индексы будут применяться для пространства произвольной
    размерности.

  \item Запятой в индексе обозначается частная производная по
    соответствующей координате ($f_{,i} := \partial_{i} f$);
    точкой с запятой --- ковариантная производная ($f_{;i} := \nabla_{i}
    f$).

  \item Для записи уравнений электродинамики в работе используется
    система СГС симметричная~\cite{sivukhin:1979:ufn::ru}.

  \end{enumerate}

\section{Гамильтонов формализм}
\label{sec:hamilton}

Можно выделить несколько вариантов гамильтонова формализма.
\begin{itemize}
\item Симплектический гамильтонов формализм~\cite{vilasi:book:hamiltonian::ru,kulyabov:2017:sfm:instantaneous_hamiltonian}.
\item Формализм Дирака--Бергмана для систем со
  связями~\cite{dirac:lectures-quantum-mechanics::ru, bergman:1955:lagrangians}.
\item Гамильтонов формализм Гамильтона--де Дондера~\cite{sardanashvily:1995:hamiltonian_formalism}.
\item Многоимпульсный гамильтонов
  формализм~\cite{sardanashvily:1995:hamiltonian_formalism,
    giachetta:1999:covariant_hamilton, giachetta:2009:advanced_classical}.
\end{itemize}

Рассмотрим общие положения гамильтонова формализма.

Пусть система описывается некоторым действием:
\begin{equation}
S[q^i] = \int \dd^4 \mathcal{L}(x^i, q^i, \Dot{q}^i).
\end{equation}
Лагранжиан (лагранжева плотность)
$(x^i, q^i, \Dot{q}^i)$ зависит от обобщённых координат $q$ и их
первых производных по времени $\Dot{q}$.

При переходе к гамильтонову формализму
система описывается обобщёнными координатами
$q^i$ и  обобщёнными импульсами:
\begin{equation}
  \label{eq:hamilton:p-L}
p_i = \pdv{\mathcal{L}}{\Dot{q}^i}.
\end{equation}

Можно построить преобразование Лежандра по скоростям:
\begin{equation}
\mathcal{H}(q^i, p_i) := p_i \Dot{q}^i − \mathcal{L}(q^i, \Dot{q}^i).
\end{equation}
Функция $\mathcal{H}$ называется гамильтонианом (гамильтоновой
плотностью). Гамильтониан системы зависит только от
обобщенных координат и импульсов. 

Используя связь между гамильтонианом и лагранжианом, рассмотрим действие:
\begin{equation}
S[q^i,p_i] = \int \dd^4 \qty(p_i\Dot{q}^i - \mathcal{H}(q^i, p_i)).
\end{equation}
Соответствующая система уравнений Эйлера--Лагранжа имеет вид:
\begin{equation}
  \label{eq:hamilton:ham-eq-S}
  \begin{gathered}
    \fdv{S}{q^i} = - \Dot{p}_i - \fdv{\mathcal{H}}{q^i} = 0,
    \\
    \fdv{S}{p_i} = \Dot{q}^i - \fdv{\mathcal{H}}{p_i} = 0.
  \end{gathered}
\end{equation}

Определим скобку Пуассона:
\begin{equation}
[f, g] := \pdv{f}{q^i} \pdv{g}{p_i} - \pdv{f}{p_i} \pdv{g}{q^i}.
\end{equation}

Тогда~\eqref{eq:hamilton:ham-eq-S} можно переписать как:
\begin{equation}
  \begin{gathered}
    \Dot{q}^i = \qty[q^i, \mathcal{H}] = \fdv{\mathcal{H}}{p_i},
    \\
    \Dot{p}_i = \qty[p_i, \mathcal{H}] = - \fdv{\mathcal{H}}{q^i}.
  \end{gathered}
\end{equation}

\section{Гамильтонова динамика со связями}
\label{sec:constrain}

Если лагранжиан вырожден по скоростям:
\begin{equation}
\det \qty[\pdv{\mathcal{L}}{\Dot{q}^i}{\Dot{q}^j}] = 0,
\end{equation}
то не возможно выразить все импульсы по
формуле~\eqref{eq:hamilton:p-L}. В этом случае
получают те импульсы, которые возможно, а для остальных вводят
понятие связи~\cite{dirac:lectures-quantum-mechanics::ru}.

Будем рассматривать систему с лагранжианом
$\mathcal{L}(x^k,q^k,p_k)$, $i=\overline{1,n}$. Также рассмотрим
набор связей:
\begin{equation}
  \varphi^a (x^k,q^k,\Dot{q}^k), \quad a=\overline{1,m},
  \quad m \leqslant n.
\end{equation}

Минимум действия, когда траектории удовлетворяют уравнениям связи,
реализуется также и на траекториях без связей, но с лагранжианом со
связями:
\begin{equation}
  L(x^k,q^k,\Dot{q}^k) =
  \mathcal{L}(x^k,q^k,\Dot{q}^k) -
  \lambda_a (x^k,q^k,\Dot{q}^k) \varphi^a (x^k,q^k,\Dot{q}^k).
\end{equation}

Уравнения Гамильтона принимают следующий вид:
\begin{equation}
  \begin{gathered}
    \Dot{q}^i = \fdv{\mathcal{H}}{p_i} + \lambda_a \pdv{\varphi^a}{p_i},
    \\
    \Dot{p}_i = - \fdv{\mathcal{H}}{q^i} - \lambda_a \pdv{\varphi^a}{q^i}.
  \end{gathered}
\end{equation}

Множители Лагранжа находятся из условия сохранения связей:
\begin{equation}
\Dot{\varphi^a} = [\varphi^a, \mathcal{H}] + \lambda_a \varphi^a = 0.
\end{equation}

\section{Гамильтониан электромагнитного поля со связями}
\label{sec:em-hamilton}

Рассмотрим построение гамильтонова формализма со связями для случая
электромагнитного поля.

Запишем лагранжиан электромагнитного поля~\cite{stratton:1948::ru}:
\begin{equation}
\mathcal{L} \qty(x^{\alpha}, A_{\beta}, A_{\alpha, \beta}) 
= 
- \frac{1}{16\pi c} F_{\alpha\beta} F^{\alpha\beta} 
\sqrt{-g}
-
\frac{1}{c^2} A_\alpha j^\alpha 
\sqrt{-g}.
\end{equation}

Поскольку $F_{00}=0$, то
\begin{equation}
\pdv[2]{\mathcal{L}}{\Dot{A}_0}=0.
\end{equation}
То есть лагранжиан нерегулярный. Следовательно, нам необходимо найти
связь для случая $p^0$. Поскольку
\begin{equation}
  p^0 = \pdv{\mathcal{L}}{\Dot{A}_0} = 0,
\end{equation}
то введём связь
\begin{equation}
  \label{eq:em-hamilton:varphi}
  \varphi := p^0 \approx 0.
\end{equation}
Символ $\approx$ обозначает, что равенство должно выполняться на
поверхности всех связей.

Запишем выражения для импульсов:
\begin{equation}
  p_i = \pdv{\mathcal{L}}{\Dot{A}^i} =
  \frac{1}{c} \pdv{\mathcal{L}}{A^i_{,0}}
  =
  -
  \frac{\sqrt{-g}}{4\pi c^2}
  \qty[
  A_{i,0} - A_{0,i}
  ]
  .
\end{equation}

Выразим производные по времени от $A_i$ через импульсы:
\begin{equation}
\Dot{A}_i = c A_{i,0} = - \frac{4\pi c^3}{\sqrt{-g}} p_i + c A_{0,i}.
\end{equation}

Построим гамильтониан:
\begin{equation}
  H =
  p^i \Dot{A}_i - \mathcal{L}
  +
  \lambda p^0
  =
  p^{i}\qty(-\frac{4\pi c^3}{\sqrt{-g}} p_i + c A_{0,i})
  +
  \frac{\sqrt{-g}}{16\pi c} F_{\alpha\beta} F^{\alpha\beta} 
  +
  \frac{\sqrt{-g}}{c^2} A_\alpha j^\alpha
  +
  \lambda p^0
\end{equation}

Уравнения Гамильтона принимают следующий вид:
\begin{gather}
  \Dot{A}_0 = \pdv{H}{p^0} = \lambda,
  \label{eq:em-hamilton:dotA0}
  \\
  \Dot{A}_{i} = \pdv{H}{p^{i}} = 
  -
  \frac{4\pi c^3}{\sqrt{-g}} p_i
  +
  c A_{0,i}
  ,
  \label{eq:em-hamilton:dotAi}
  \\
  \Dot{p}^0 = -\fdv{H}{A_0} =
  - \pdv{H}{A_0}
  + \partial_{\alpha} \pdv{H}{A_{0,\alpha}}
  =
  -c p^{i}_{,i} + \frac{\sqrt{-g}}{c^2} j^0,
  \label{eq:em-hamilton:dotp0}
  \\
  \Dot{p}^i = - \fdv{H}{A_i} = - \pdv{H}{A_i}
  + \partial_{\alpha} \pdv{H}{A_{i,\alpha}}
  =
  - \frac{\sqrt{-g}}{c^2} j^{i}
  +
  \frac{\sqrt{-g}}{16\pi c} F^{\alpha i}_{,\alpha}
  .
  \label{eq:em-hamilton:dotpi}
\end{gather}

Покажем, что полученная система уравнений эквивалентна уравнениям
Максвелла (что представляется очевидным).

Из уравнения~\eqref{eq:em-hamilton:dotAi} получим
\begin{equation}
  \label{eq:em-hamilton:dotAi:pi}
  p_i =
  \frac{\sqrt{-g}}{4\pi c^3}
  \qty(
  c A_{0,i} - \Dot{A}_{i} 
  )
  =
  \frac{\sqrt{-g}}{4\pi c^2}
  \qty(
  A_{0,i} - A_{i,0}
  )
  =
  \frac{\sqrt{-g}}{4\pi c^2}
  F_{i 0}.
\end{equation}
Поднимем индексы в уравнении~\eqref{eq:em-hamilton:dotAi:pi} и подставим
в уравнение~\eqref{eq:em-hamilton:dotpi}.
Записав компоненты тензора $F^{{\alpha}{\beta}}$
\begin{equation}
F^{{\alpha}{\beta}}=
\begin{pmatrix}
0 & -{D}^1 & -{D}^2 & -{D}^3 \\
{D}^1 & 0 & -{H}_3 & {H}_2 \\
{D}^2 & {H}_3 & 0 & -{H}_1 \\
{D}^3 & -{H}_2 & {H}_1 & 0
\end{pmatrix},
\label{eq:g^ab}
\end{equation}
получим уравнение
\begin{equation}
  \frac{1}{\sqrt{{}^{3}g}}
  \qty[\partial_{{j}}H_{{k}}-\partial_{{k}}H_{{j}}]
  =
  -\frac{1}{c}\partial_t D^{{i}}+\frac{4\pi}{c}j^{{i}}.
\end{equation}

Аналогично из~\eqref{eq:em-hamilton:dotp0} с учётом
уравнения~\eqref{eq:em-hamilton:dotAi:pi} и
связи~\eqref{eq:em-hamilton:varphi}
получим ещё одно уравнение Максвелла.
\begin{equation}
\frac{1}{\sqrt{{}^{3}g}}
\partial_{{i}} \qty(\sqrt{{}^{3}g} D^{{i}}) = 4\pi \rho.
\end{equation}

Таким образом, мы показали, что полученный гамильтониан даёт
неоднородные уравнения Максвелла. При этом очевидно, что однородные уравнения
Максвелла получаются не из гамильтониана, а из тождества Бьянки для
тензора $F_{\alpha \beta}$.

\section{Заключение}
\label{sec:conclusion}

Авторы применили методику Дирака--Бергмана к уравнениям Максвелла. Полученные уравнения
совпадают с неоднородными уравнениями Максвелла. Особенностью
рассмотрения является то, что рассматривались уравнения Максвелла  в
ковариантной записи в произвольных римановых координатах.

\def\acknowledgmentsname{Благодарности}

\begin{acknowledgments}

Публикация подготовлена при поддержке Программы РУДН «5-100»
и при финансовой поддержке РФФИ в рамках научного проекта
\textnumero~19-01-00645.

\end{acknowledgments}

  \ifthenelse{\boolean{xetex}}
  {\bibliographystyle{ugost2008l}}
  {\bibliographystyle{gost2008l}}

\bibliography{bib/hamilton-geom/cite}